\documentclass[apj,numberedappendix]{emulateapj}
\usepackage{apjfonts}

\usepackage{lscape}
\usepackage{epsf}

\shorttitle{GCMF IN THE SOMBRERO GALAXY}
\shortauthors{CHANDAR, FALL, \& MCLAUGHLIN}
\slugcomment{ApJ Letters, in press}

\begin{document}

\title{DENSITY DEPENDENCE OF THE MASS FUNCTION OF
        GLOBULAR STAR CLUSTERS IN THE SOMBRERO GALAXY
        AND ITS DYNAMICAL IMPLICATIONS}

\author{Rupali Chandar      \altaffilmark{1,2},
         S. Michael Fall     \altaffilmark{3}, and
         Dean E. McLaughlin  \altaffilmark{4}
         }

\altaffiltext{1}{Carnegie Observatories, 813 Santa Barbara Street,
         Pasadena, CA 91101}
\altaffiltext{2}{current address: Department of Physics \& Astronomy,
         University of Toledo, Toledo, OH 43606; Rupali.Chandar@utoledo.edu}
\altaffiltext{3}{Space Telescope Science Institute, 3700 San
         Martin Drive, Baltimore, MD 21218; fall@stsci.edu}
\altaffiltext{4}{School of Physical and Geographical Sciences,
         Keele University, Keele ST5 5BG, United Kingdom;
         dem@astro.keele.ac.uk}

\begin{abstract}

We have constructed the mass function of globular 
star clusters in the Sombrero galaxy in bins of 
different internal half-mass density $\rho_h$ 
and projected galactocentric distance $R$.
This is based on the published measurements of the 
magnitudes and effective radii of the clusters by
Spitler et al. (2006) in $BVR$ images taken with 
the ACS on {\it HST}.
We find that the peak of the mass function $M_p$
increases with $\rho_h$ by a factor of about 4 but 
remains nearly constant with $R$.
Our results are almost identical to those presented
recently by McLaughlin \& Fall (2007) for globular 
clusters in the Milky Way.
The mass functions in both galaxies agree with a 
simple, approximate model in which the clusters 
form with a Schechter initial mass function and
evolve subsequently by stellar escape driven by
internal two-body relaxation. 
These findings therefore undermine recent claims that 
the present peak of the mass function of globular
clusters must have been built into the initial 
conditions.

\end{abstract}


\section{INTRODUCTION}

One of the most fundamental properties of a system of
star clusters is its mass function, defined here such
that $\psi(M)dM$ is the number of clusters with masses
between $M$ and $M+dM$ [i.e., $\psi(M) \equiv dN/dM$].
There is currently a debate about whether the shape of
the mass function of old globular clusters primarily
reflects dynamical evolution (Fall \& Zhang 2001) or
initial conditions (Vesperini et al. 2003).
A key observation here is that the turnover or 
peak of $\psi(M)$ at $M_p \approx (1-2)\times 10^5 
M_{\odot}$ is nearly ``universal,'' varying little 
among galaxies of different masses and types and 
from one location to another within galaxies 
(Harris 2001).
Interpreted naively, these facts tend to favor the
scenario in which the present shape of $\psi(M)$
was imprinted when the clusters formed or were
relatively young.
On the other hand, the observed shape $\psi(M) 
\approx {\rm const}$ below $M_p$ is a clear signature 
of long-term dynamical evolution driven by internal 
two-body relaxation (Fall \& Zhang 2001; Waters et al. 
2006; Jord{\'a}n et al. 2007).

Another signature of relaxation-driven evolution,
recently emphasized by McLaughlin \& Fall (2007,
hereafter MF07), is that the peak mass should be
higher in subsamples of clusters with greater
internal densities and thus higher evaporation
rates.
The expected dependence is roughly $M_p \propto
\rho_h^{1/2}$, where $\rho_h = 3M/8{\pi}r_h^3$
is the mean density within the median (i.e.,
half-mass) radius $r_h$.
MF07 found essentially this dependence of $M_p$ 
on $\rho_h$ for the old globular clusters in the 
Milky Way, and they argued that this, together 
with the observation $\psi(M) \approx {\rm const}$ 
for $M \la M_p$, provided compelling evidence for 
the scenario in which dynamical evolution determines 
the present shape of $\psi(M)$.

In this Letter, we report on a similar study of the
old globular clusters in the Sombrero galaxy (M104,
NGC 4594).
Our goal is to check whether the MF07 findings apply
only to the clusters in the Milky Way or are
representative of the clusters in other galaxies.
We use the published measurements by Spitler et 
al. (2006) of the luminosities and effective radii 
of the globular clusters in images taken with the 
Advanced Camera for Surveys (ACS) on the {\it Hubble 
Space Telescope (HST)} as part of the Hubble Heritage 
Program (GO-9714).
The Sombrero galaxy is ideal for this project because
it has a large retinue of clusters ($N > 600$),
nearly all them are resolved in the ACS images, and
there is relatively little contamination by the disk.

\begin{figure*}
\centerline{\hfil
   \rotatebox{0}{\includegraphics[height=95mm,width=150mm]{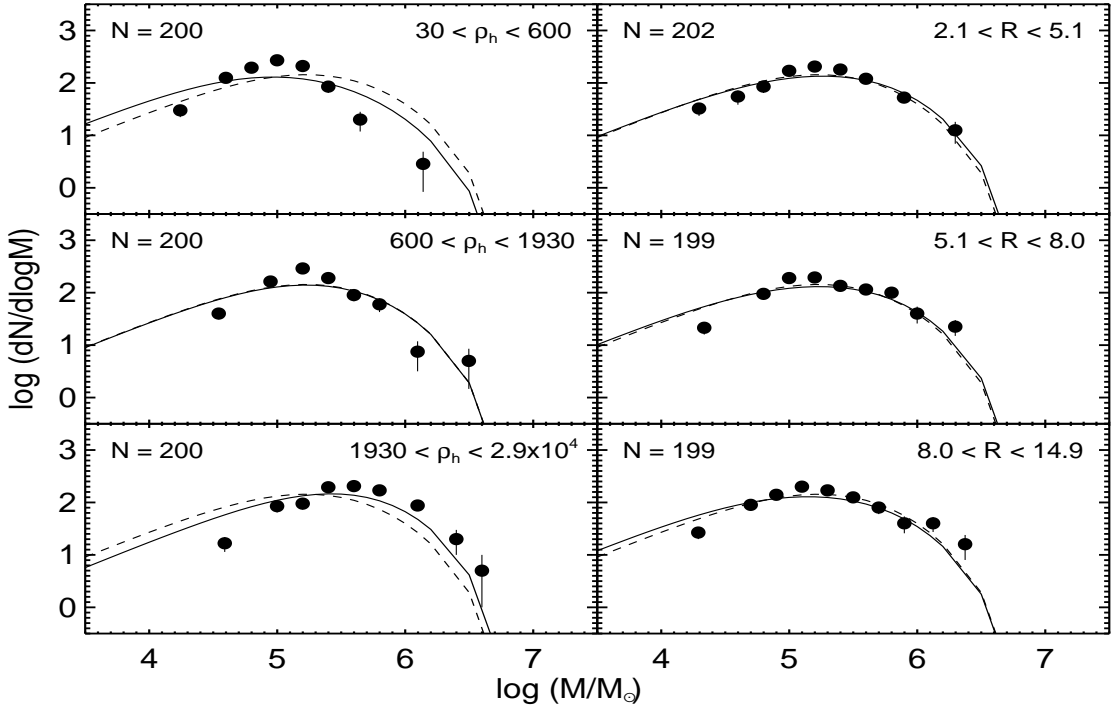}}
\hfil}
\caption{Mass function of globular clusters in the 
Sombrero galaxy. 
The panels on the left-hand side show three ranges 
of internal half-mass density $\rho_h$ in units of 
$M_{\odot}~{\rm pc}^{-3}$, as indicated (excluding 
clusters with $R < 2.1$~kpc). 
The panels on the right-hand side show three 
ranges of projected galactocentric distance $R$ 
in kpc, as indicated (excluding clusters with 
$\rho_h < 30~M_{\odot}~{\rm pc}^{-3}$). 
The vertical bars represent Poisson ($\sqrt{N}$) 
errors.
The solid curves were computed from equation~(1) 
as described in the text by summing over terms
with individual values of $\rho_{h,i}$ for all 
clusters in the $\rho_h$ and $R$ bins specified 
in each panel. 
The dashed curves, which are the same in all 
six panels, were computed from equation~(1) with 
a single term with the median value of $\rho_h$ 
for the full sample of 600 clusters.}
\end{figure*}

\section{OBSERVATIONS}

The ACS images are arranged in a $3 \times 2$ mosaic
with overall angular dimensions of $600'' \times 400''$,
corresponding to $26~{\rm kpc} \times 17~{\rm kpc}$ at
the adopted 9.0~Mpc distance of the Sombrero galaxy.
The images are moderately deep; for each pointing, 
four integrations of 675~s, 500~s, and 350~s were 
made with the F435W ($B$), F555W ($V$), and F625W 
($R$) filters, respectively.
Spitler et al. (2006) selected candidate globular 
clusters based on the following criteria:
{\em (i)} integrated colors in the ranges $0.9< B-R 
<1.7$ and $0.5 < B-V < 1.1$; 
{\em (ii)} apparent $V$ magnitude brighter than 24.3 
(roughly the 95\% completeness limit);
{\em (iii)} angular extent at least 0.3~pixels wider 
than the point-spread function (PSF);
{\em (iv)} ellipticity less than 0.5;
{\em (v)} location outside the dust lane; and
{\em (vi)} visual appearance more like a globular 
cluster than a star or galaxy.
This produced a sample of 659 candidate globular 
clusters that is believed to be nearly complete 
brighter than $M_V \approx -5.5$ and nearly free 
from contamination by foreground stars and 
background galaxies. 

Spitler et al. (2006) present the magnitudes, 
effective (i.e., half-light) radii $r_e$, and 
projected galactocentric distances $R$ of all 
the clusters in their sample.\footnote{The 
quantities denoted here by $r_e$ and $R$ are
denoted by $R_{\rm hl}$ and $R_{\rm GC}$ in
Table~2 of Spitler et al. (2006).}
We estimate the mass of each cluster from $M = 
L_V(M/L_V)$, with the total $V$-band luminosity 
$L_V$ (after correction for Galactic interstellar 
extinction) and the adopted mass-to-light ratio 
$M/L_V = 1.5~M_{\odot}/L_{\odot}$ (a typical value 
for old globular clusters in the Milky Way; see 
McLaughlin 2000). 
We estimate the internal half-mass density from the 
definition $\rho_h \equiv 3M/8{\pi}r_h^3$ and the
standard 3D-2D conversion, $r_h = (4/3)r_e$
(neglecting any internal mass/light segregation; 
see Spitzer 1987).

Figure~1 shows the main results of this paper: 
the mass function of the globular clusters in the 
Sombrero galaxy in three bins of internal half-mass 
density $\rho_h$ (left-hand panels) and three bins 
of projected galactocentric distance $R$ 
(right-hand panels).
We plot $dN/d{\log}M$ rather than $dN/dM$ 
for ease of comparison with the more familiar 
magnitude distributions.
Based on the tests described below, we exclude 
the 59 clusters with either 
$\rho_h < 3 0~M_{\odot}~{\rm pc}^{-3}$ or 
$R < 0.8'$ (2.1~kpc), because the sample becomes
less complete at low internal densities (for any 
$R$) and small galactocentric distances (for any 
$\rho_h$). 
The mass functions plotted here show the familiar
single-peaked shape. 
What is striking, however, is that the peak mass 
increases from $M_p \approx 1 \times 10^5~M_{\odot}$ 
in the low-$\rho_h$ bin, to $M_p \approx 2 \times 
10^5~M_{\odot}$ in the middle-$\rho_h$ bin, to $M_p 
\approx 4 \times 10^5~M_{\odot}$ in the 
high-$\rho_h$ bin.
At the same time, the peak mass remains roughly
constant at $M_p \approx 2 \times 10^5~M_{\odot}$
for all three $R$ bins.
Both of these results have high statistical
significance and are similar to the MF07 findings 
for the globular clusters in the Milky Way (see 
their Fig.~2).

We have replotted Figure~1 with many different 
binnings in $M$, $\rho_h$, and $R$, and always 
recover the same result---that $M_p$ increases
with $\rho_h$ but is nearly constant with 
$R$---to within the statistical uncertainties. 
We have also made diagrams like Figure~1 for
subsamples of clusters divided by color at 
$B - R = 1.3$ and find only minor differences.
This is relevant because the globular clusters 
in the Sombrero galaxy, like those in many other 
galaxies, have a bimodal distribution of colors.
It is believed that the red, metal-rich clusters 
($B - R > 1.3$) are somewhat younger than the 
blue, metal-poor clusters ($B - R < 1.3$).
The spectra of several red and blue clusters
indicate, however, that they are all old, with
ages that differ by less than 3~Gyr from each 
other and from the ages of globular clusters 
in the Milky Way (Larsen et al. 2002).

We have also made independent measurements of 
the magnitudes and effective radii of all the 
clusters in the Spitler et al. (2006) sample.
Our procedure is similar to theirs, with only
two minor differences: our measurements of $V$
include $r_e$-dependent aperture corrections and
our measurements of $r_e$ are from the $V$-band
images alone. 
Both sets of measurements are generally in good
agreement, the mean offset (us minus Spitler et 
al.) and RMS scatter being only 0.06 and 0.03 
in $V$ and 0.04 and 0.07 in $\log r_e$.
These small deviations are at the levels 
expected for the slightly different procedures.
As another check, we have inserted artificial 
clusters with specified properties into the 
$V$-band images and measured them in the same 
way as the real clusters.
These experiments indicate that the uncertainties
in $V$ and $r_e$ increase, while the completeness
of the sample decreases, for diffuse clusters 
($\rho_h \la 30~M_{\odot}~{\rm pc}^{-3}$) and
those near the galactic center ($R < 0.8'$).
We have imposed these restrictions on $\rho_h$ 
and $R$ in most of our analysis (as noted above), 
but we have confirmed that the relations between
$M_p$, $\rho_h$, and $R$ are nearly identical
in the full Spitler et al. sample.
Finally, we have repeated the entire analysis
using only our measurements of $V$ and $r_e$, 
and find results that are practically 
indistinguishable from those based on 
the Spitler et al. measurements. 
These checks give us confidence in our claim 
that $M_p$ has a significant dependence on 
$\rho_h$ but not $R$.

\section{MODEL}

\begin{figure}
\centerline{\hfil
   \rotatebox{0}{\includegraphics[width=90mm]{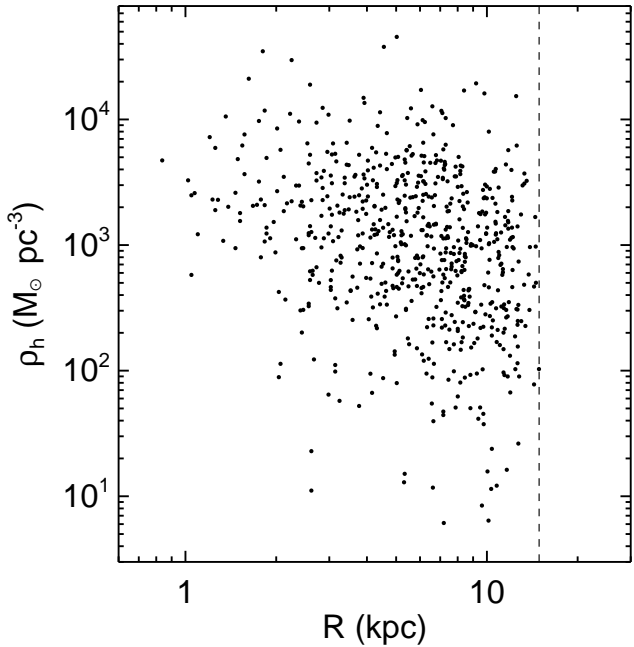}}
\hfil}
\caption{Scatter plot of internal half-mass 
density $\rho_h$ against projected galactocentric 
distance $R$ for all clusters in the Spitler (2006) 
et al. sample.
The vertical dashed line shows the maximum value
of $R$ accessible in the $3 \times 2$ mosaic of 
ACS images.} 
\end{figure}

In this section, we explore the implications of
the results displayed in Figure 1.
As we have already noted, the peak of the mass
function $M_p$ is a factor of about 4 higher 
in the bottom left-hand panel than in the top
left-hand panel, while the median internal 
density $\rho_h$ within these subsamples 
increases by a factor of about 13.
This is close to the expected scaling 
$M_p \propto \rho_h^{1/2}$ for relaxation-driven
evolution, although the comparison is necessarily
crude because the range of $\rho_h$ included in
each of the panels is large.
The puzzle of why $M_p$ has little or no 
dependence on projected galactocentric distance 
$R$ is explained by Figure~2, a plot of $\rho_h$ 
against $R$ for all clusters in the sample. 
Evidently, there is a huge scatter and only a 
mild (anti)correlation between $\rho_h$ and $R$,
thus accounting nicely for the weak dependence 
of $M_p$ on $R$ despite the relatively strong 
dependence of $M_p$ on $\rho_h$.
(See Fig.~1 of MF07 for the analogous situation 
in the Milky Way.)

To make a more precise and informative comparison,
we follow the procedure developed by MF07, which in
turn is based on the simple (``classical'') model
for the escape of stars from tidally limited clusters
driven by internal relaxation alone (see Section 3.2
of Spitzer 1987 and references therein). In this
model, the mass of each cluster decreases linearly
with time, $M(t) = M_0 - \mu_{\rm ev}t$, where $M_0$
is the initial mass and $\mu_{\rm ev} = -dM/dt \propto
M/t_{\rm rh} \propto \rho_h^{1/2}$ is the constant
rate of mass loss ($t_{\rm rh}$ being the half-mass
relaxation time).
For a population of clusters with the same internal
density $\rho_h$ and age $\tau$, the present and
initial mass functions are related by $\psi(M)
=\psi_0(M + \mu_{\rm ev}\tau)$ (Fall \& Zhang 2001).
This implies that the present mass function will
have a peak at $M_p \approx \mu_{\rm ev}\tau$, a
low-mass shape $\psi(M) \approx {\rm const}$ (for
$M \la M_p$), and a high-mass shape $\psi(M)
\approx \psi_0(M)$ (for $M \ga M_p$).

For the initial mass function, we adopt the
Schechter (1976) distribution, $\psi_0(M) \propto
M^{\beta} \exp(-M/M_c)$, with a power-law exponent
$\beta = -2$ and an exponential cutoff at $M_c = 1
\times 10^6 M_{\odot}$.
This choice is motivated by the observed power-law
shape of the mass function of young clusters in the
Antennae galaxies, $\psi_0(M) \propto M^{-2}$ for
$M \la 1 \times 10^6 M_{\odot}$ (Zhang \& Fall
1999), and the fact that the mass functions of old
globular clusters in many galaxies decrease more
rapidly than a power law for $M \ga 1 \times 10^6
M_{\odot}$ (Burkert \& Smith 2000; Jord{\'a}n et 
al. 2007).
We assume this mass function holds at some 
``initial'' time, near the formation epoch but 
after a brief period of ``infant mortality'' caused 
by the activity of massive stars (photoionization, 
winds, jets, supernovae), since we are concerned 
here only with the long-term dynamical evolution
of the clusters. 
In our model, this evolution flattens much of the 
power-law part of $\psi_0(M)$ (for $M \la M_p$) 
but preserves the exponential part.
Thus, fortunately, most of our results are not 
sensitive to our assumed initial conditions.

The mass function for a population of clusters
with a range of internal densities is just the
sum of the mass functions for the clusters of
each density.
From the relations above, we obtain a formula
equivalent to equation (3) of MF07:
\begin{equation}
\psi(M) = \sum_i A_i (M + \Delta_i)^{-2}
           \exp[-(M + \Delta_i)/M_c].
\end{equation}
Here, $\Delta_i = (\mu_{\rm ev}\tau)_i 
\propto \rho_{h,i}^{1/2}$ is the mass 
lost over the lifetime of a cluster, 
$A_i = \Delta_i/E_2(\Delta_i/M_c)$ is the
required normalization constant for each term
($E_2$ being an exponential integral), and the
sum is over all clusters in the population.
This model is based on the same dynamical evolution 
and initial conditions as the Fall-Zhang (2001) 
model, but the internal densities of clusters are 
no longer coupled to their orbits and hence to 
assumptions about the space and time dependence 
of the galactic potential.
Thus, equation (1) is equally valid for spherical 
and triaxial, static and evolving potentials.

We have computed $dN/d\log{M} = (M/\log{e})
\psi(M)$ from equation~(1) and plotted the 
results in Figure~1.
The coefficient $B$ in the mass-loss formula
\begin{equation}
\Delta_i = B 
(\rho_{h,i}/10^3 M_{\odot}~{\rm pc}^{-3})^{1/2}
10^5 M_{\odot}
\end{equation}
was determined as follows.
We first fitted a model with a single term in
equation~(1)---an ``evolved Schechter function'' 
in the nomenclature of Jord{\'a}n et al. 
(2007)---to the mass function for the full sample 
of 600 clusters.
The minimum $\chi^2$ occurs for $\Delta=2.4 \times 
10^5~M_{\odot}$ and hence $\mu_{\rm ev} = 1.8 \times
10^{-5}~M_{\odot}~{\rm yr}^{-1}$ for $\tau = 13$~Gyr.
Associating this rate of mass loss with the median 
internal density of clusters in the full sample, 
$\rho_h = 1.1 \times 10^3~M_{\odot}~{\rm pc}^{-3}$, 
gives $B = 2.3$. 
The resulting ``average'' mass function for the
globular clusters in the Sombrero galaxy is plotted
as the dashed curve in each of the six panels of 
Figure~1.
This is very similar to the corresponding function 
derived by the same procedure by MF07 for the globular 
clusters in the Milky Way. 

We next computed the mass functions for the three 
$\rho_h$ bins and three $R$ bins from equation~(1)
with a separate term for each cluster.
In doing so, we used the same coefficient $B = 2.3$
derived above for the whole sample to relate the
individual values of $\Delta_i$ and $\rho_{h,i}$.
The results are plotted as the solid curves in 
Figure~1.
These provide good (although not perfect)
representations of the observed mass functions, 
with generally similar shapes and peaks at about 
the right masses.
In particular, the model $M_p$ increases by a 
factor of about 3 between the low-$\rho_h$ and 
high-$\rho_h$ bins, while remaining almost 
constant for the different $R$ bins.
The first of these relations is slightly 
weaker than $M_p \propto \rho_h^{1/2}$ because 
of the exponential cutoff at $M_c = 1 \times 
10^6~M_{\odot}$.\footnote
{The simple scaling $M_p \propto \rho_h^{1/2}$ 
is exact in the limit $M_c \gg M_p$ (MF07).}

The observed mass function is slightly narrower 
or more sharply peaked than the model curves, 
especially in the left-hand panels of Figure 1.
One factor that might contribute to this is 
residual incompleteness for faint (i.e., low-mass) 
clusters, particularly in the
lowest-density bin.  
Another is our simplifying assumption that all
clusters have the same mass-to-light ratio.  
In reality, there will be
a distribution of $M/L_V$, and this will broaden 
the mass function somewhat relative to the luminosity 
function [which we have adopted as
a proxy for $\psi(M)$].  
A more speculative possibility that would
further improve the fits is that the cutoff $M_c$ 
in the initial mass function might increase slightly 
with $\rho_h$ instead of remaining
fixed as we have assumed here.  
In any case, it is clear from Figure~1
that our simple, relaxation-driven model captures the essential
features of the observed mass function even without these adjustments.

\section{DISCUSSION}

Our findings for the globular clusters in 
the Sombrero galaxy corroborate those of MF07 
for the globular clusters in the Milky Way.
The Sombrero sample is larger ($N = 600$ vs 
$N=146$) but, because of the modest sensitivity
to extended sources and the restricted angular
extent of the {\it HST} observations, it covers
smaller ranges of internal density 
($6 \la \rho_h \la 5 \times 10^4 
M_{\odot}~{\rm pc}^{-3}$ vs $3 \times 
10^{-2} \la \rho_h \la 6 \times 10^4 
M_{\odot}~{\rm pc}^{-3}$) and galactocentric
distance ($2 \la R \la 15$~kpc vs $0.6 \la
R \la 123$~kpc).
In both galaxies, the peak of the mass function 
increases with internal density---approximately 
as $M_p \propto \rho_h^{1/2}$---but remains 
nearly constant with galactocentric distance.
These relations are consistent with each other 
because $\rho_h$ correlates only weakly with $R$.
Thus, we now see that whether the peak mass 
appears ``universal'' depends crucially on 
one's perspective. 
In terms of galactocentric distance, it may 
be; but in terms of internal density, the more 
physically meaningful quantity, it certainly is 
not.

We have interpreted these observations in the 
context of a simple, approximate model in which 
clusters form with a Schechter initial mass 
function and are subsequently disrupted by 
stellar escape driven by internal two-body 
relaxation.
Our model predicts present mass functions 
similar to the observed ones, including the
correct dependence on both $\rho_h$ and $R$.
We compute mass-loss rates and hence $\psi(M)$ 
directly from the observed $\rho_h$, and only 
then do we use the observed $\rho_h$-- $R$ 
distribution to reexpress $\psi(M)$ in 
terms of $R$.
Our approach completely avoids having to
specify the $\rho_h$--$R$ distribution in
the past and thus sidesteps the interesting 
but difficult problem of explaining physically 
how the present $\rho_h$--$R$ distribution came 
about (which ultimately will require a detailed 
understanding of the formation and evolution 
of both galaxies and star clusters).
Instead, we assume only that the present  
$\rho_h$ of extant clusters are good guides 
to their past $\rho_h$ and thus to their 
average mass-loss rates over a Hubble time.
For a more complete discussion of these issues,
we refer the interested reader to MF07.

Most previous models link the internal densities 
and hence the mass-loss rates of clusters to their 
orbits in static, spherical galactic potentials 
(adopted for analytical simplicity). 
To account for the observed weak radial gradient
in the mass functions of globular clusters, these 
models are forced to have orbital distributions 
with strong radial anisotropy at large 
galactocentric distances, possibly more than 
is allowed by observations in the Milky Way (Fall 
\& Zhang 2001) and certainly more than allowed in 
M87 (Vesperini et al. 2003).
However, as Fall \& Zhang (2001) have emphasized,
these conclusions are primarily a consequence of
the simplifying assumptions about the galactic
potential, not the underlying premise about the 
disruption of clusters. 
In more realistic models, galaxies form and 
evolve by a hierarchical series of collisions, 
mergers, and other accretion events.
Galactic potentials in this case are time-dependent 
and non-spherical, and the orbital energies and 
angular momenta of clusters are not conserved. 
The orbits and positions of clusters are 
rearranged many times and any initial gradients 
in $M_p$ or $\rho_h$ are inevitably weakened if 
not eliminated, consistent with the observations 
plotted in Figures~1 and 2.
By comparison, the idealized models with static, 
spherical potentials overpredict gradients in 
$M_p$ and $\rho_h$ and therefore represent a 
limiting (i.e., ``worst-case'') scenario.
\footnote{The strong radial mixing that 
occurs within evolving galaxies is readily 
apparent in movies generated from simulations 
of cosmological structure formation, although, 
to our knowledge, this has not been quantified 
in terms of radial transition probabilities 
or rates.}

Our results also have an important bearing 
on recent attempts to infer the initial form 
of the mass function of globular clusters.
Vesperini et al. (2003) have advocated
a gaussian-like or lognormal initial mass 
function with a built-in peak near $10^5~ 
M_{\odot}$ to account for the observed shape
and weak radial variation of the present 
mass function in the context of models 
with static, spherical galactic potentials
(see also Vesperini 2000, 2001).
As noted above, this is not the only, or even
the most natural solution of this problem.
Our model provides a direct counterexample 
to these claims: an initial mass function with 
a power-law shape (for $M \la 10^6 M_{\odot}$) 
evolves into the present mass function with a 
peak at $M_p \approx 2 \times 10^5 M_{\odot}$ 
and with the observed dependence of $M_p$  
on $\rho_h$ and $R$.

\acknowledgements 
We thank Forrest Hamilton for creating the
ACS mosaic images used in this project. 
Our work is supported in part by NASA grants 
AR-09539.1-A and GO-10402.05-A.  
$HST$ data are obtained at STScI, which is operated 
by AURA, Inc., under NASA contract NAS5-26555.

\end{document}